\documentclass{ws-procs9x6}
\usepackage{graphicx}

\begin{document}

\title{Pentaquark baryon production in nuclear reactions
\footnote{\uppercase{S}upported in part by the \uppercase{US}
\uppercase{N}ational \uppercase{S}cience \uppercase{F}oundation under
\uppercase{G}rant \uppercase{N}o. \uppercase{PHY}-0098805 and the 
\uppercase{W}elch \uppercase{F}oundation under \uppercase{G}rant 
\uppercase{N}o. \uppercase{A}-1358.}}

\author{C.~M. Ko and W. Liu}

\address{Cyclotron Institute and Physics Department\\
Texas A\&M University\\
College Station, TX 77843-3366, USA\\
E-mail: ko@comp.tamu.edu, weiliu@neo.tamu.edu}

\maketitle

\abstracts{Using a hadronic model with empirical coupling constants 
and form factors, we have evaluated the cross sections for
the production of exotic pentaquark $\Theta^+$ and/or 
$\Xi_5^+$ and $\Xi_5^{--}$ in reactions induced by photons, nucleons,
pions, and kaons on nucleon targets. We have also predicted the
$\Theta^+$ yield in relativistic heavy ion collisions using a kinetic
model that takes into account both $\Theta^+$ production from the
coalescence of quarks and antiquarks in the quark-gluon plasma and the
effects due to subsequent hadronic absorption and regeneration.}

\section{Introduction}

One of the most exciting recent experimental results in hadron
spectroscopy is the narrow baryon state that was inferred
from the invariant mass spectrum of $K^+n$ or $K^0p$ in nuclear
reactions induced by photons \cite{nakano}, kaons \cite{barmin}, and 
protons \cite{cosy}. The extracted mass of about 1.54 GeV and width of
less than 21-25 MeV are consistent with those of the pentaquark baryon
$\Theta^+$ consisting of $uudd\bar s$ quarks and predicted in the
chiral soliton model \cite{diakonov}. However, the reported large
width is limited by experimental resolutions as the actual width is
expected to be much smaller \cite{arndt}. Its existence has
also been verified in the Skyrme model \cite{prasz}, the chiral quark
model \cite{hosaka}, the constituent quark model \cite{stancu}, the QCD sum
rules \cite{zhu}, and the lattice QCD \cite{csikor}.  Although most
models predict that $\Theta^+$ has spin 1/2 and isospin 0, their
predictions on $\Theta^+$ parity vary widely.  While the soliton model
gives a positive parity and the lattice QCD studies favor a negative
parity, the quark model can give either positive or negative parities, 
depending on whether quarks are correlated or not.  To help determine
the quantum numbers of $\Theta^+$, studies have been carried out to 
understand its production mechanism in these reactions \cite{oh,nam}. 
Using a hadronic model with empirical coupling constants and form
factors, we have evaluated the cross sections for $\Theta^+$
production from these reactions \cite{liu1,liu2,liu4} and also those
for other multistrange exotic pentaquark baryons 
$\Xi_5^{+}(uussd\bar d)$ and $\Xi_5^{--}(ddssu \bar u)$
\cite{liu3}. The latter has been observed in proton-proton collisions
at center-of-mass energy of 17.2 GeV by the NA49 Collaboration at SPS
\cite{alt}, with mass 1.86 GeV and width about 18 MeV due to detector
resolution. Moreover, the yield of $\Theta^+$ in heavy ion collisions
at the Relativistic Heavy Ion Collider (RHIC) has been studied using
the quark coalescence model \cite{chen}. In this talk, we report the
results from our studies.

\section{$\Theta^+$ production in photonucleon reactions}

\begin{figure}[ht]
\centerline{\includegraphics[height=2cm,width=8cm,angle=0]{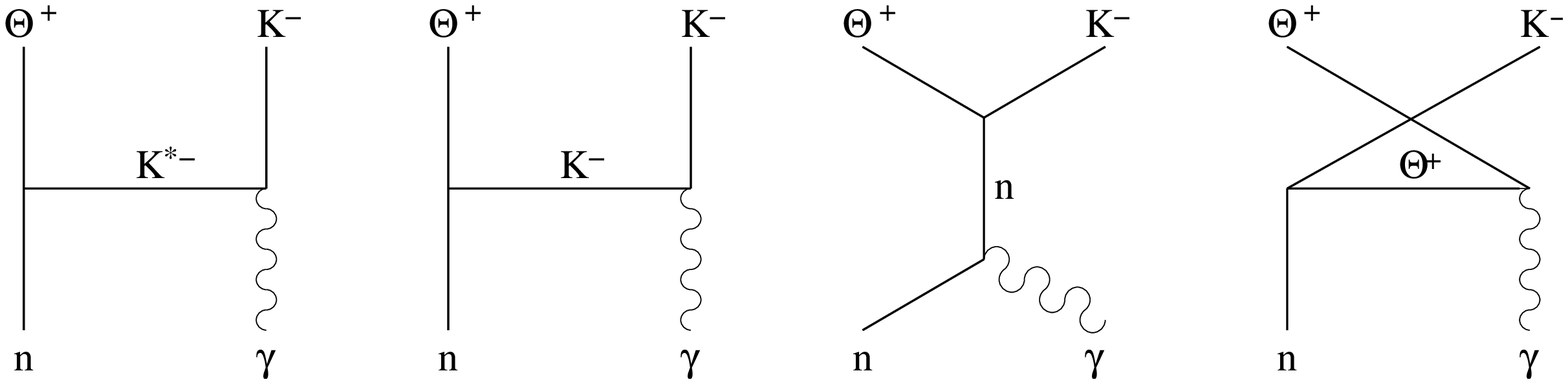}}
\caption{Diagrams for $\Theta^+$ production from the reaction
$\gamma n\to K^-\Theta^+$.}
\label{diagramg}
\end{figure}

For $\Theta^+$ production from photonucleon reactions, possible 
processes are $\gamma p\to\bar K^0\Theta^+$, $\gamma n\to
K^-\Theta^+$, $\gamma p\to\bar K^{*0}\Theta^+$, and $\gamma n\to
K^{*-}\Theta^+$. As an illustration, we show in Fig.\ref{diagramg}
four possible diagrams for the reaction $\gamma n\to K^-\Theta^+$, involving
the $t$-channel $K$- and $K^*$-exchange, the $s$-channel nucleon pole, and
the $u$-channel $\Theta^+$-exchange. To evaluate their amplitudes, 
we need the photon coupling to $\Theta^+$ as well as the coupling 
constants $g_{NK\Theta}$ and $g_{NK^*\Theta}$, besides the well-known 
couplings of photon to $K^-$, $K^{*-}$, and neutron. Since the
anomalous magnetic moment of $\Theta^+$ is not known empirically and
theoretical estimates give a much smaller value than that of nucleons
\cite{nam,kim}, we have neglected its contribution and included only the
coupling of photon to the charge of $\Theta^+$. For the coupling
constant $g_{KN\Theta}$, it is proportional to the square root
of $\Theta^+$ width and depends on the spin and parity of $\Theta^+$. 
Taking the $\Theta^+$ width to be $1$ MeV, which is consistent
with available $K^+N$ and $Kd$ data \cite{arndt}, we then have 
$g_{KN\Theta}\approx 1$ if $\Theta^+$ has spin 1/2 and positive
parity. It is reduced to $g_{KN\Theta}\approx 0.14$ if the $\Theta^+$
parity is negative due to the absence of a centrifugal barrier
between nucleon and kaon. For the coupling constant $g_{K^*N\Theta}$,
it would have been $g_{K^*N\Theta}\approx 0.6 g_{KN\Theta}$ if
$N(1710)$, which has same partial decay width of $~15$ MeV to $N\pi$
and $N\rho$, were to belong to the same antidecuplet as $\Theta^+$. On
the other hand, it is given by $g_{K^*N\Theta}=\sqrt{3}g_{KN\Theta}$
in the fall-apart decay model \cite{carlson}. In our work, we have
assumed that $g_{K^*N\Theta}$ is zero or has the same magnitude as 
$g_{KN\Theta}$.

\begin{figure}[ht]
\centerline{
\includegraphics[height=3.3cm,width=4cm,angle=-90]{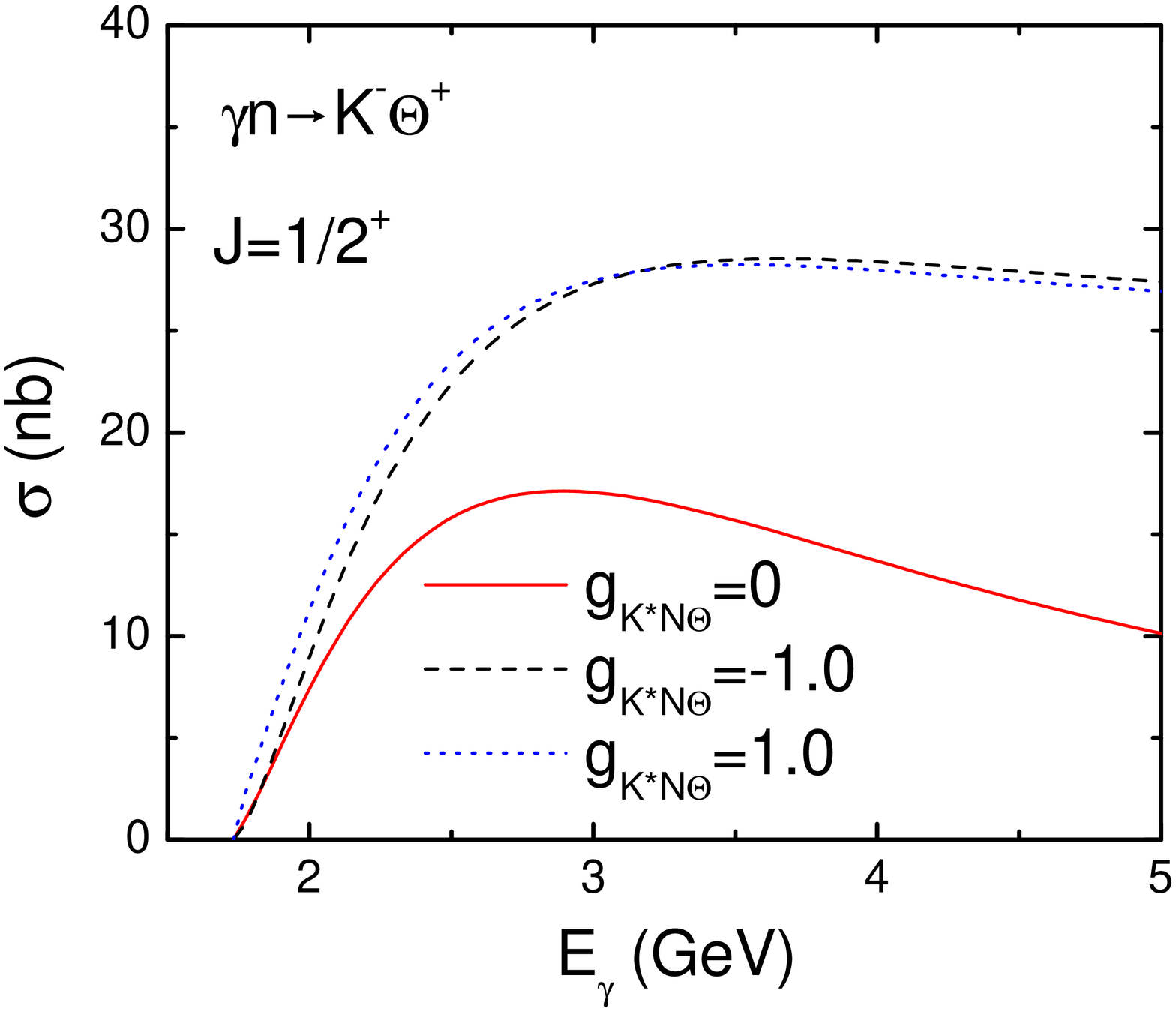}
\hspace{-0.77cm}
\includegraphics[height=3.3cm,width=3.9cm,angle=-90]{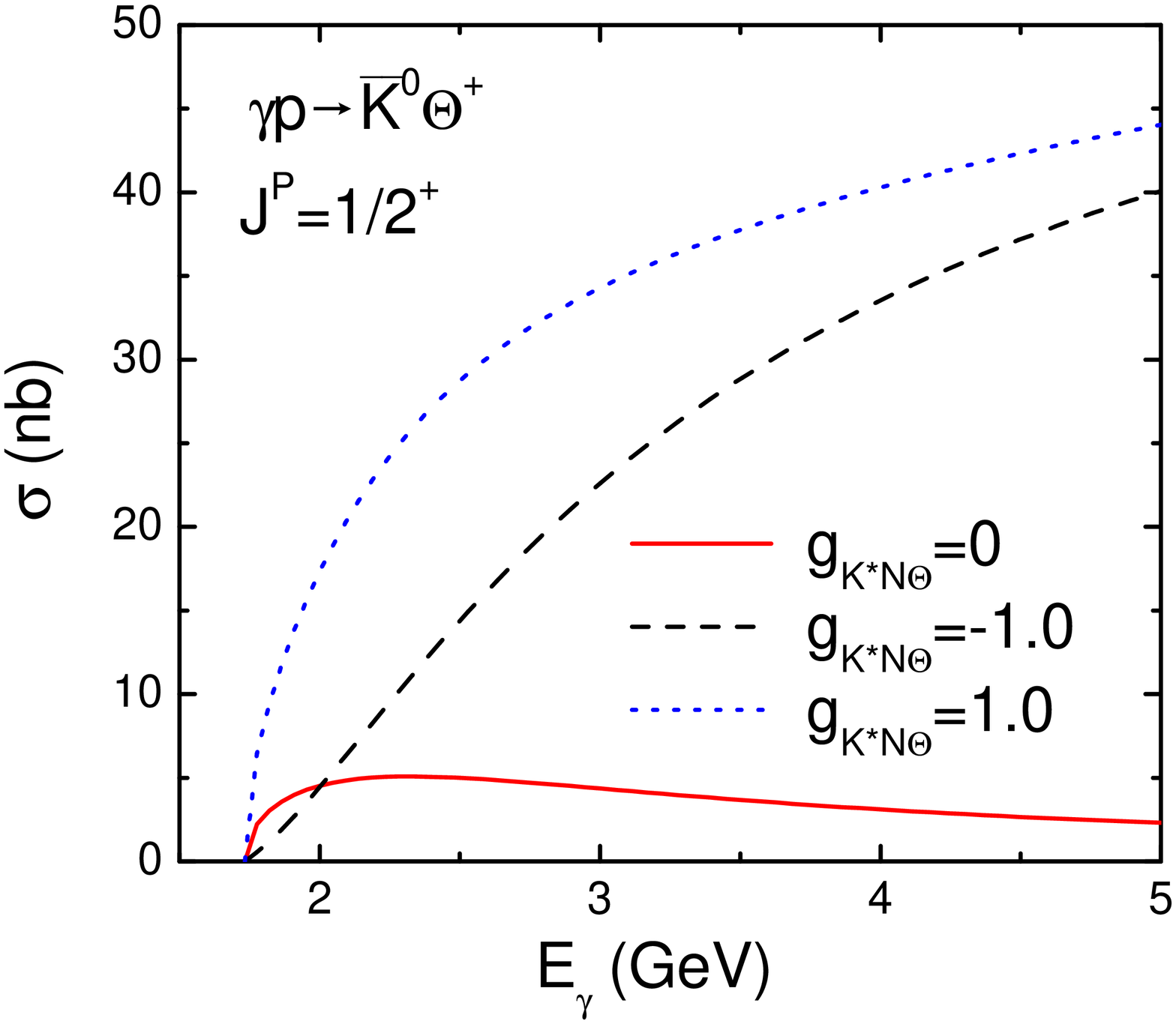}
\hspace{-0.77cm}
\includegraphics[height=3.3cm,width=4cm,angle=-90]{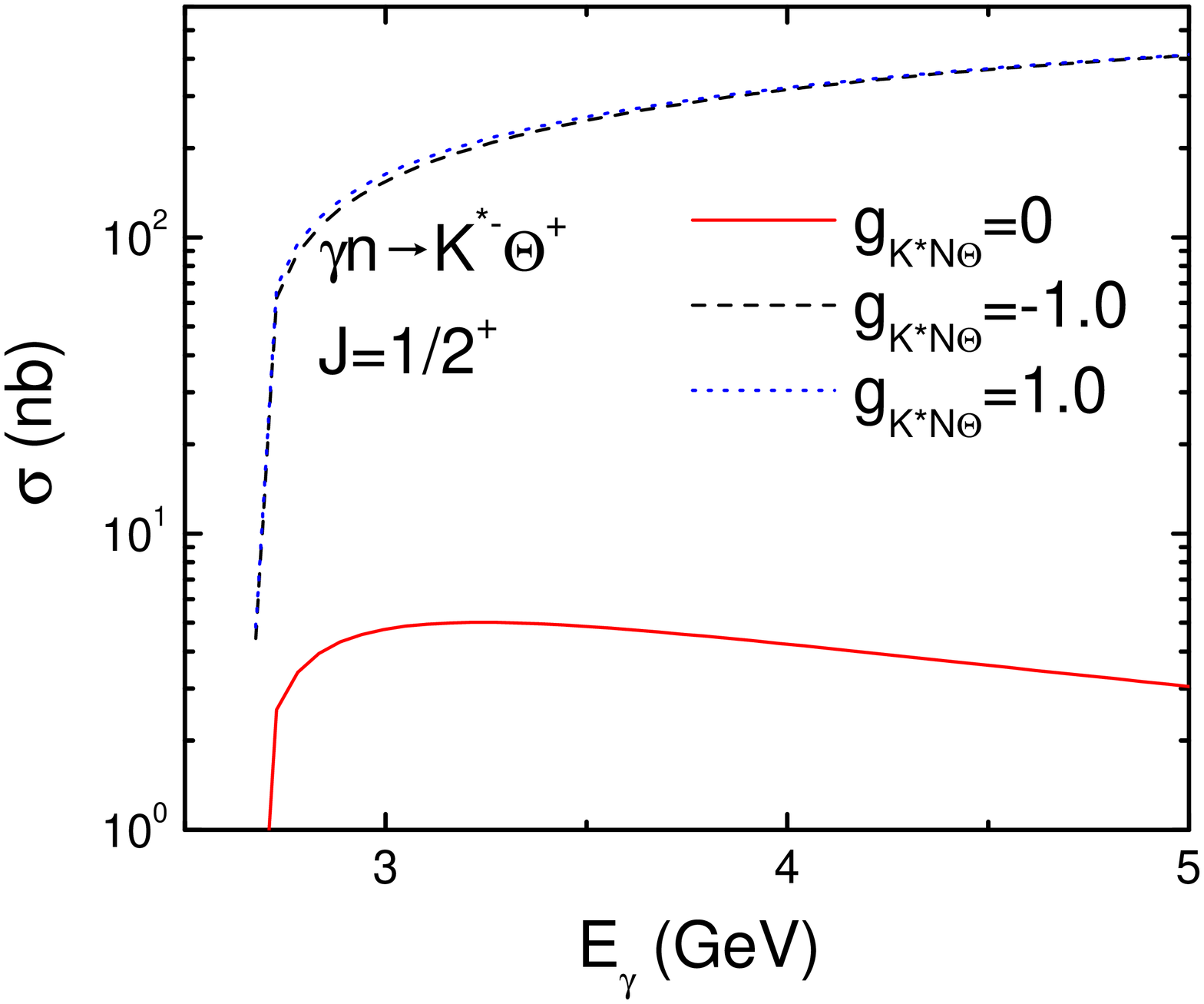}
\hspace{-0.77cm}
\includegraphics[height=3.3cm,width=4cm,angle=-90]{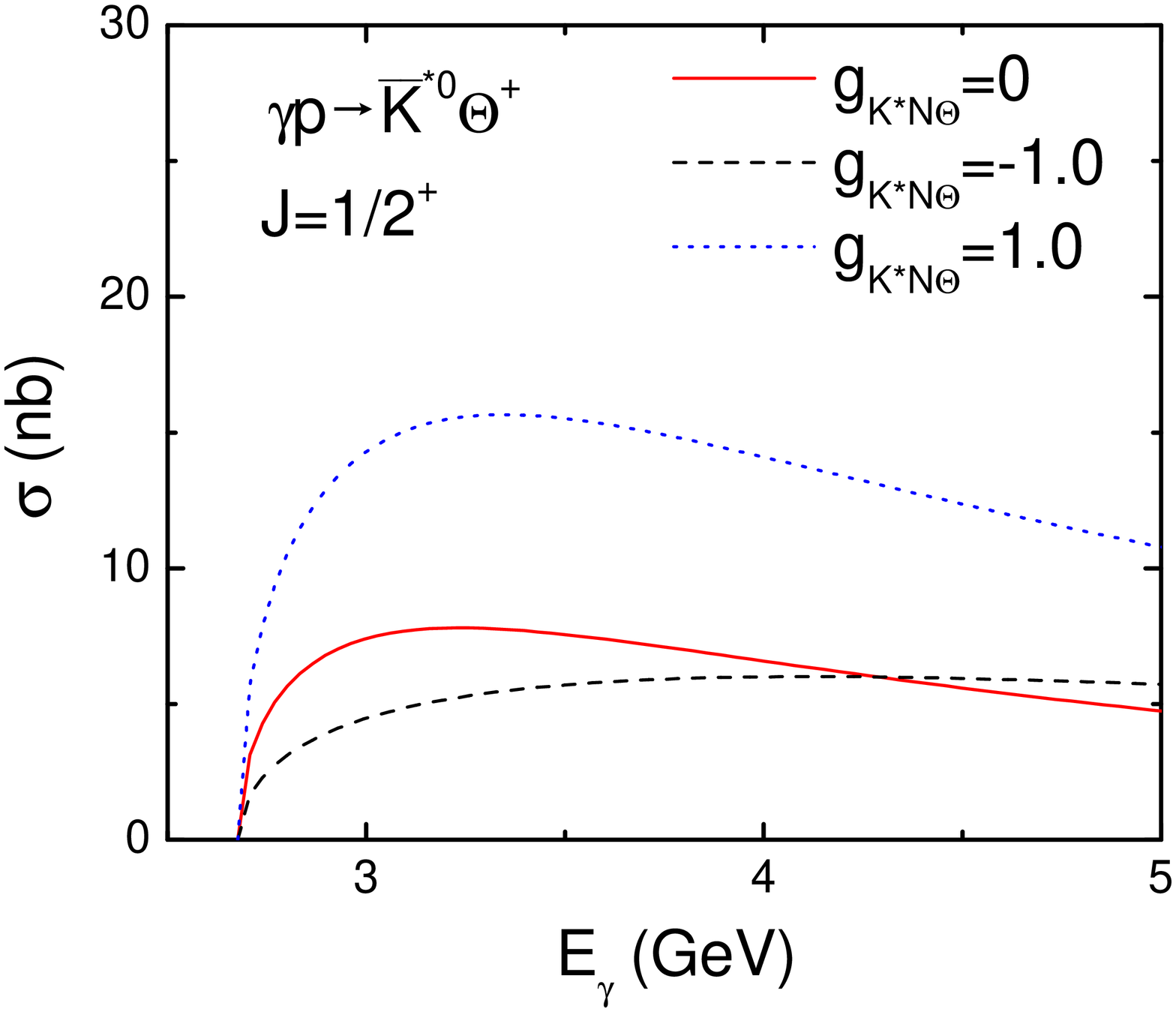}}
\centerline{
\includegraphics[height=3.3cm,width=4cm,angle=-90]{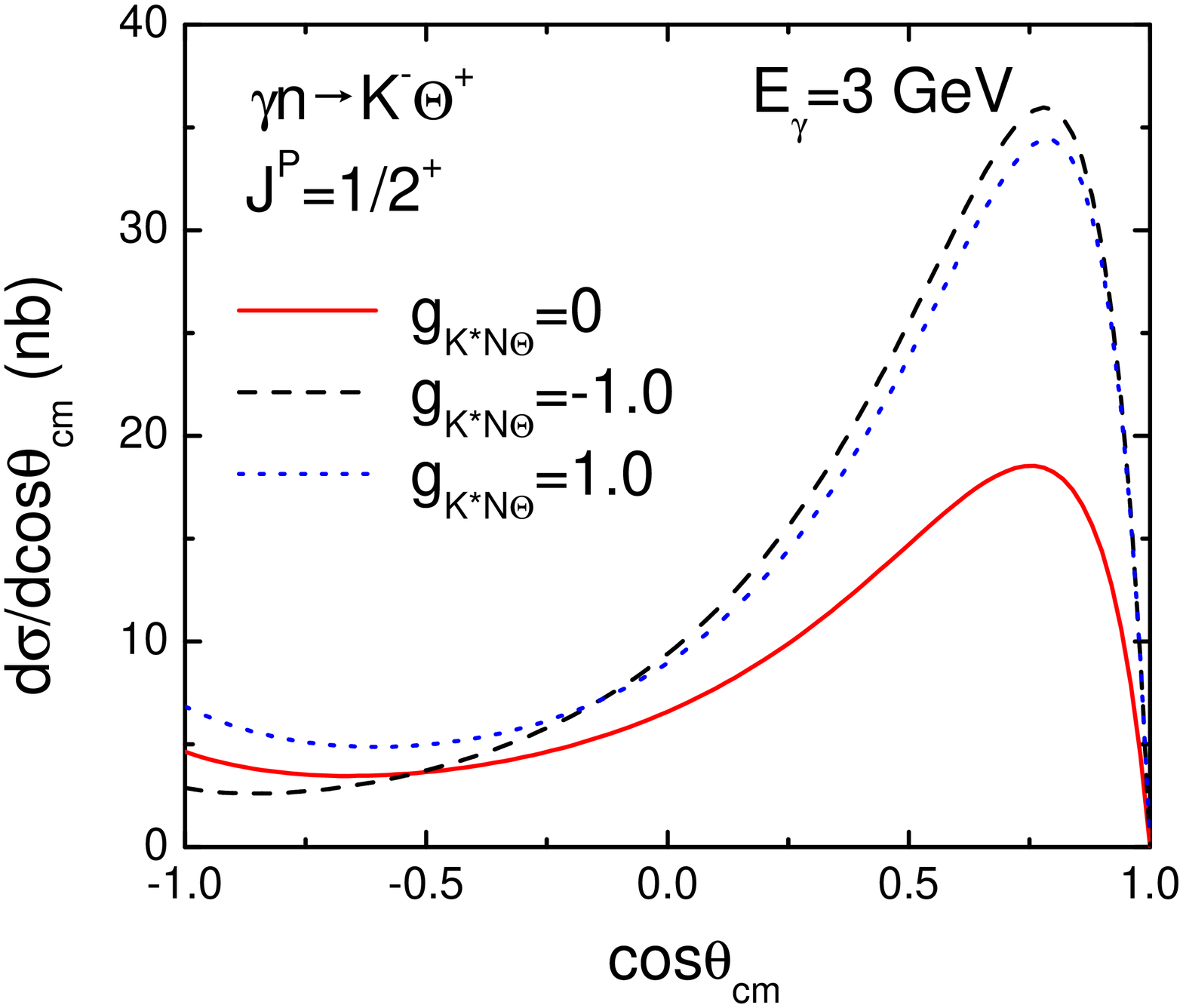}
\hspace{-0.77cm}
\includegraphics[height=3.3cm,width=4cm,angle=-90]{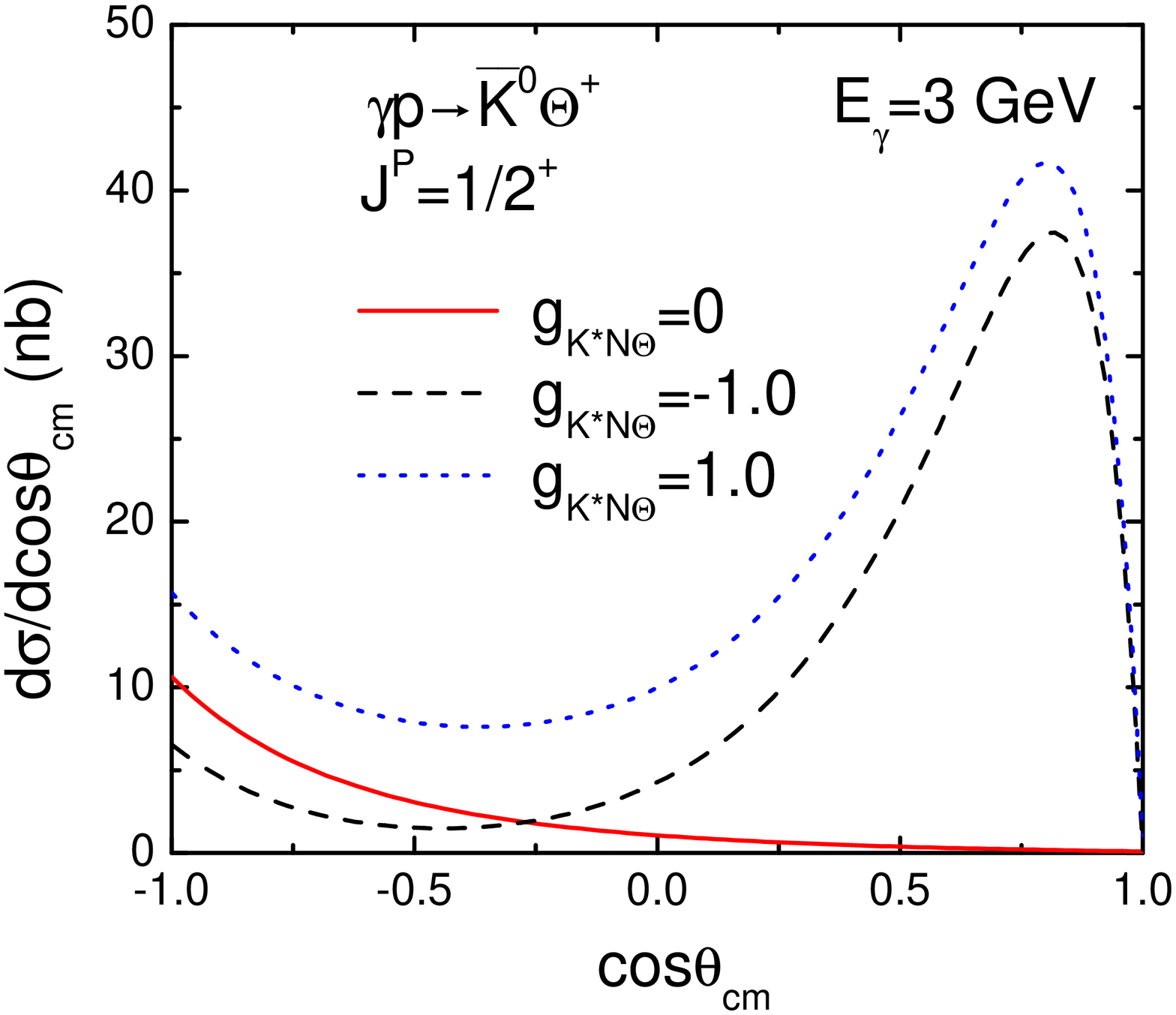}
\hspace{-0.77cm}
\includegraphics[height=3.3cm,width=4cm,angle=-90]{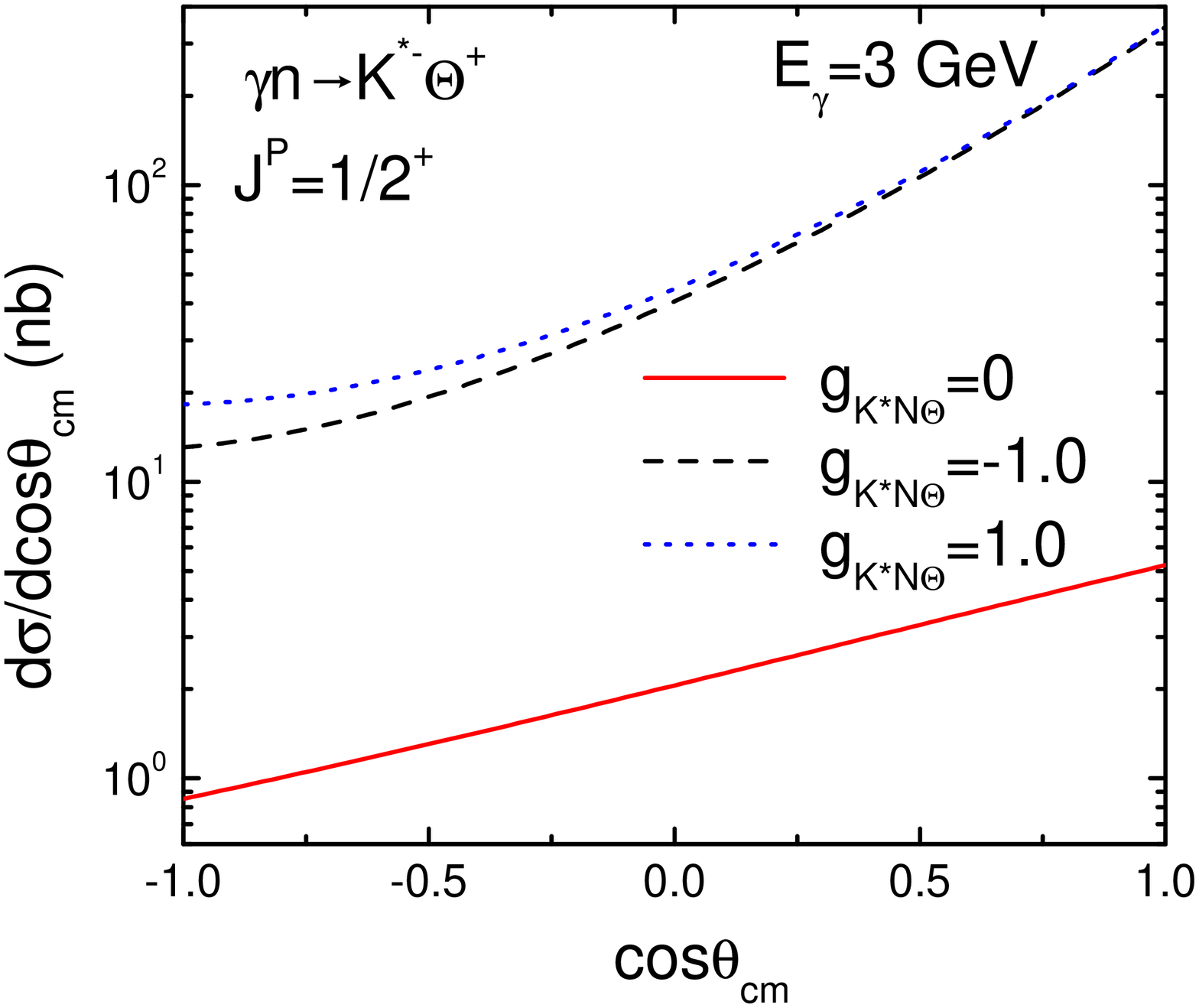}
\hspace{-0.77cm}
\includegraphics[height=3.3cm,width=4cm,angle=-90]{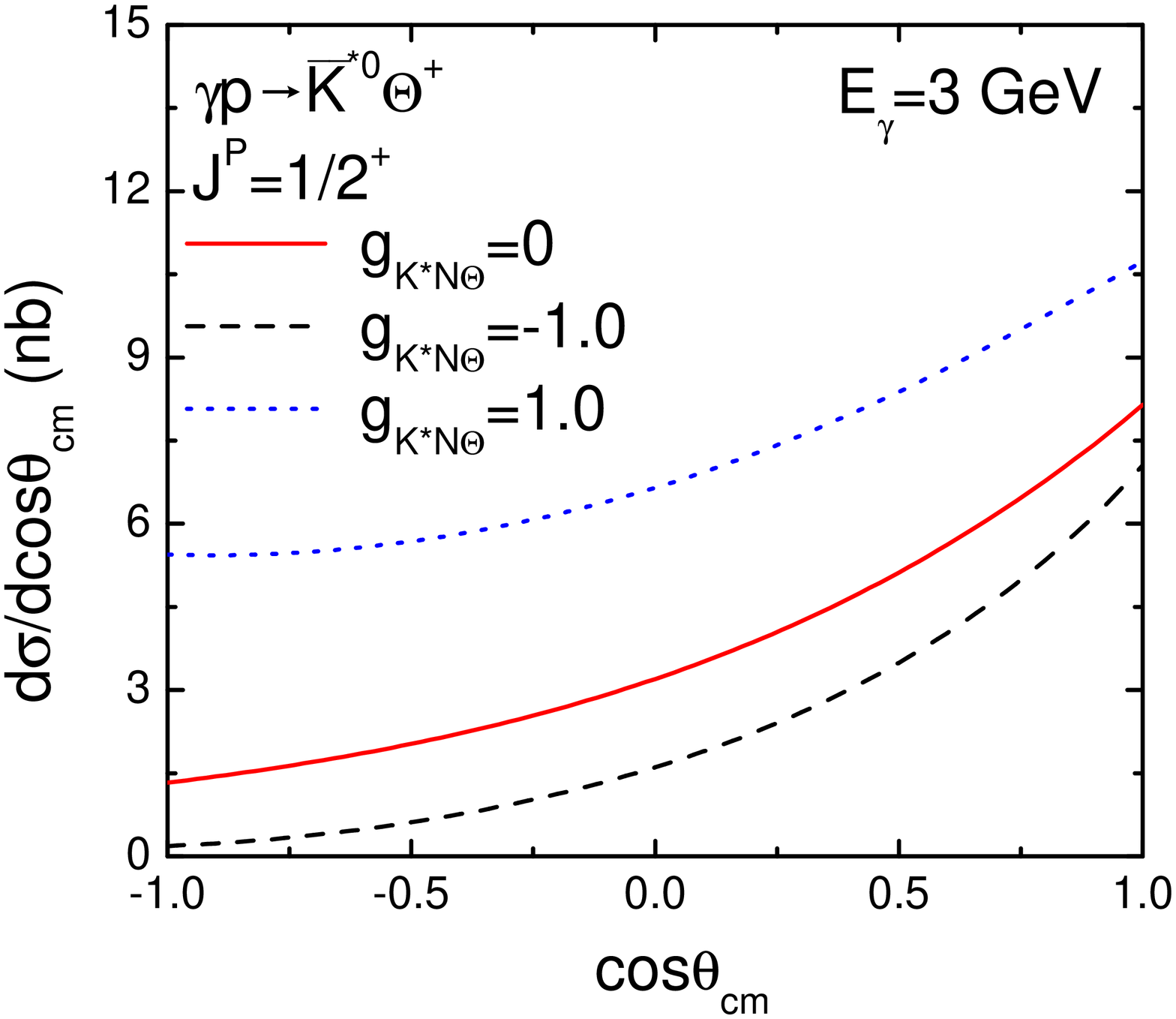}}
\caption{Total (upper panels) and differential (lower panels) cross
sections for $\Theta^+$ production from photo-nucleon reactions.} 
\label{crossg}
\end{figure}

To take into account the finite sizes of hadrons, we have included
for each amplitude a form factor of the form 
$F(x)=\Lambda^4/[\Lambda^4+(x-m_x^2)^2]$, where $x=s$,
$t$, and $u$ with corresponding masses $m_x=m_N$, $m_K$ or
$m_{K^*}$, and $m_\Theta$ of the off-shell particles at strong
interaction vertices \cite{davidson}.  After restoring the gauge
invariance of resulting total amplitude by adding a suitable
contact term in the interaction Lagrangian, the cross section for the
reaction $\gamma n\to K^-\Theta^+$ has been evaluated in
Ref. \cite{liu4} with a cutoff parameter $\Lambda\approx 1.2$ GeV,
that is determined from fitting the measured cross section for charmed
hadron production from photon-proton reactions with two-body final
states \cite{liu5}. The resulting total and differential cross
sections for the reaction $\gamma n\to K^-\Theta^+$ are shown in the
leftmost panels of Fig.\ref{crossg}. The cross sections at photon
energy $E_\gamma=3$ GeV are about 30 nb and 15 nb with and without
$K^*$ exchange, respectively, and the produced $\Theta^+$ peaks at
forward direction in the center-of-mass system. For negative parity
$\Theta^+$, the cross section is reduced by about a factor of 10,
although the coupling constant $g_{KN\Theta}$ is about a factor of 7
smaller than that for positive parity $\Theta^+$, and the angular
distribution remains forward peaked. 
 
The cross sections for $\Theta^+$ production from other photon-nucleon
reactions $\gamma p\to\bar K^0\Theta^+$, $\gamma n\to
K^{*-}\Theta^+$, and $\gamma p\to\bar K^{*0}\Theta^+$
can be similarly evaluated, and their cross sections are shown in
other panels of Fig.\ref{crossg}. The cross section of about 35 nb
for the reaction $\gamma p\to\bar K^0\Theta^+$ at $E_\gamma=3$ GeV is
comparable to the revised value of about 50 nb measured in SAPHIR
experiment at Bonn Electron Stretcher Accelerator.

\section{$\Theta^+$ production in proton-proton reactions}

\begin{figure}[ht]
\centerline{
\includegraphics[height=4cm,width=4cm,angle=0]{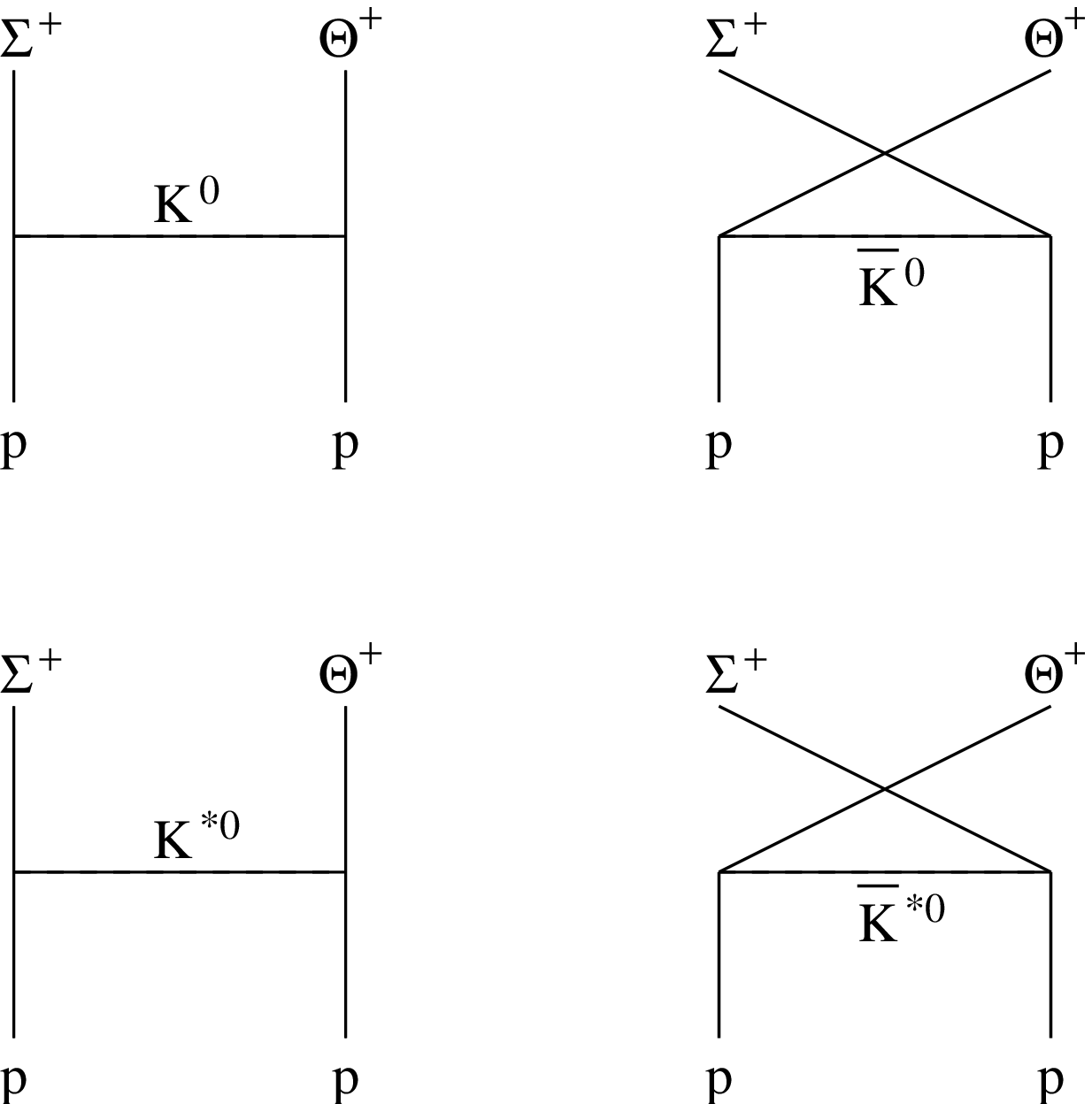}
\hspace{1cm}
\includegraphics[height=4cm,width=4cm,angle=0]{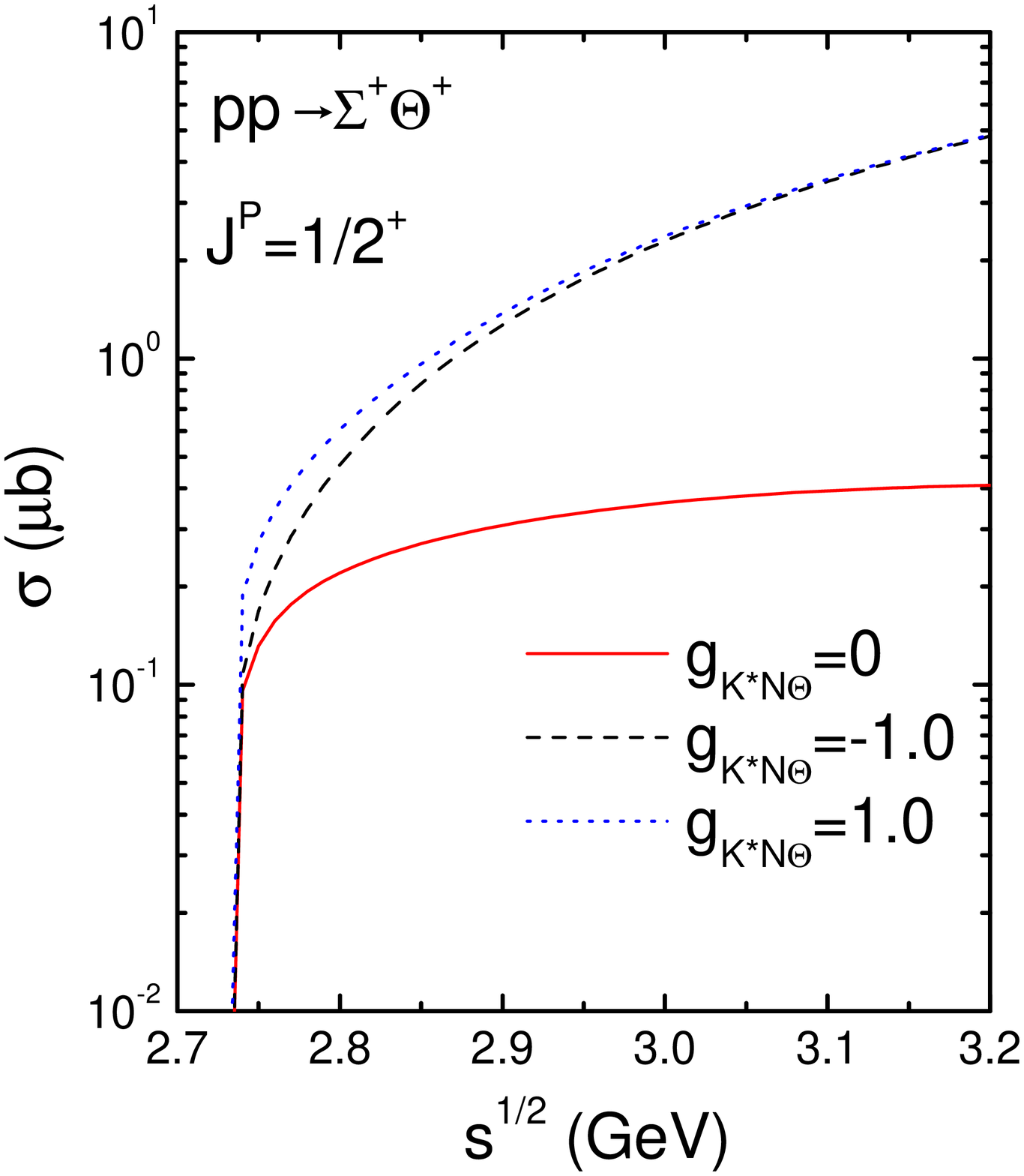}}
\caption{Diagrams (left panels) and cross sections (right panel) for
$\Theta^+$ production from the reaction $pp\to \Sigma^+\Theta^+$.}
\label{diagramp} 
\end{figure}

For $\Theta^+$ production in proton-proton reactions, possible
diagrams are shown in the left panels of Fig.\ref{diagramp} involving both
$t$- and $u$-channel exchange of $K$ and $K^*$. The coupling constants 
$g_{KN\Sigma}\approx -3.78$ and $g_{K^*N\Sigma}=3.25$ are obtained
from well-known $g_{\pi NN}=13.5$ and $g_{\rho NN}=3.25$ by SU(3)
relations. The form factors at interaction vertices are taken to be 
$F({\bf q}^2)=\Lambda^2/(\Lambda^2+{\bf q}^2)$, with ${\bf q}$ being
the three momentum of exchanged $K$ or $K^*$. Using a cutoff parameter 
$\Lambda=0.42$ GeV, obtained from fitting the experimental cross
sections for the reactions $pp\to K^+\Lambda p$ and 
$pp\to D^+\Lambda_c p$ with similar hadronic models \cite{liu6},
the resulting cross section for the reaction $pp\to\Sigma^+\Theta^+$
is shown in the right panel of Fig.\ref{diagramp} for $g_{KN\Theta}=1$
and different values of $g_{K^*N\Theta}$. The value of 0.3 $\mu$b
obtained with $g_{K^*N\Theta}=1.0$ at center-of-mass energy
$\sqrt{s}=2.75$ GeV, corresponding to a beam momentum $p_{\rm
beam}=2.95$ GeV/c, is comparable to the 0.4 $\mu$b measured in the
COSY experiment \cite{cosy}. 

\section{$\Theta^+$ production in pion- and kaon-nucleon reactions}

\begin{figure}[ht]
\centerline{\includegraphics[height=2cm,width=8cm,angle=0]{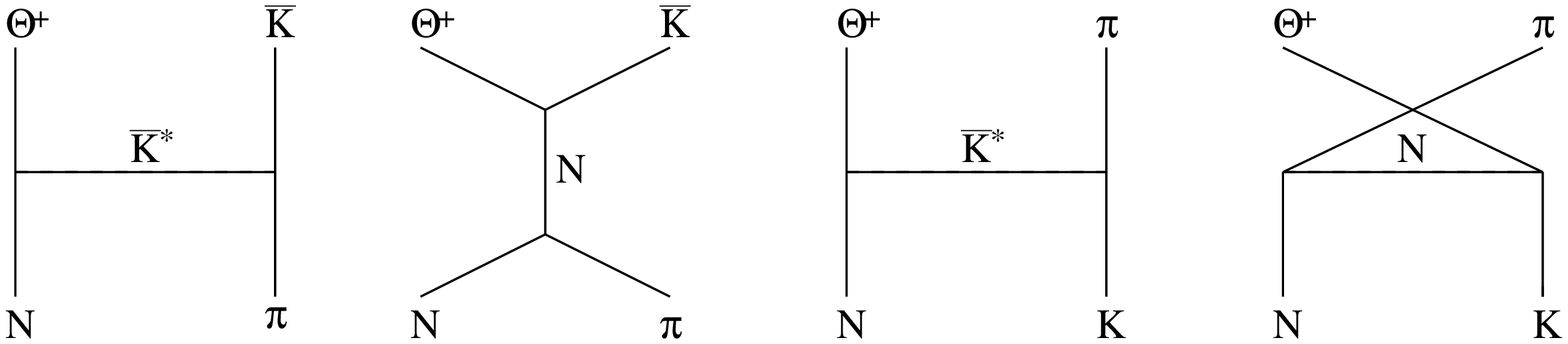}}
\caption{Diagrams for $\Theta^+$ production from the reactions
$\pi N\to\bar K\Theta^+$ (left two panels) and $KN\to\pi\Theta^+$
(right two panels).} 
\label{diagrampik}
\end{figure}

\begin{figure}[ht]
\centerline{
\includegraphics[height=3.3cm,width=4cm,angle=-90]{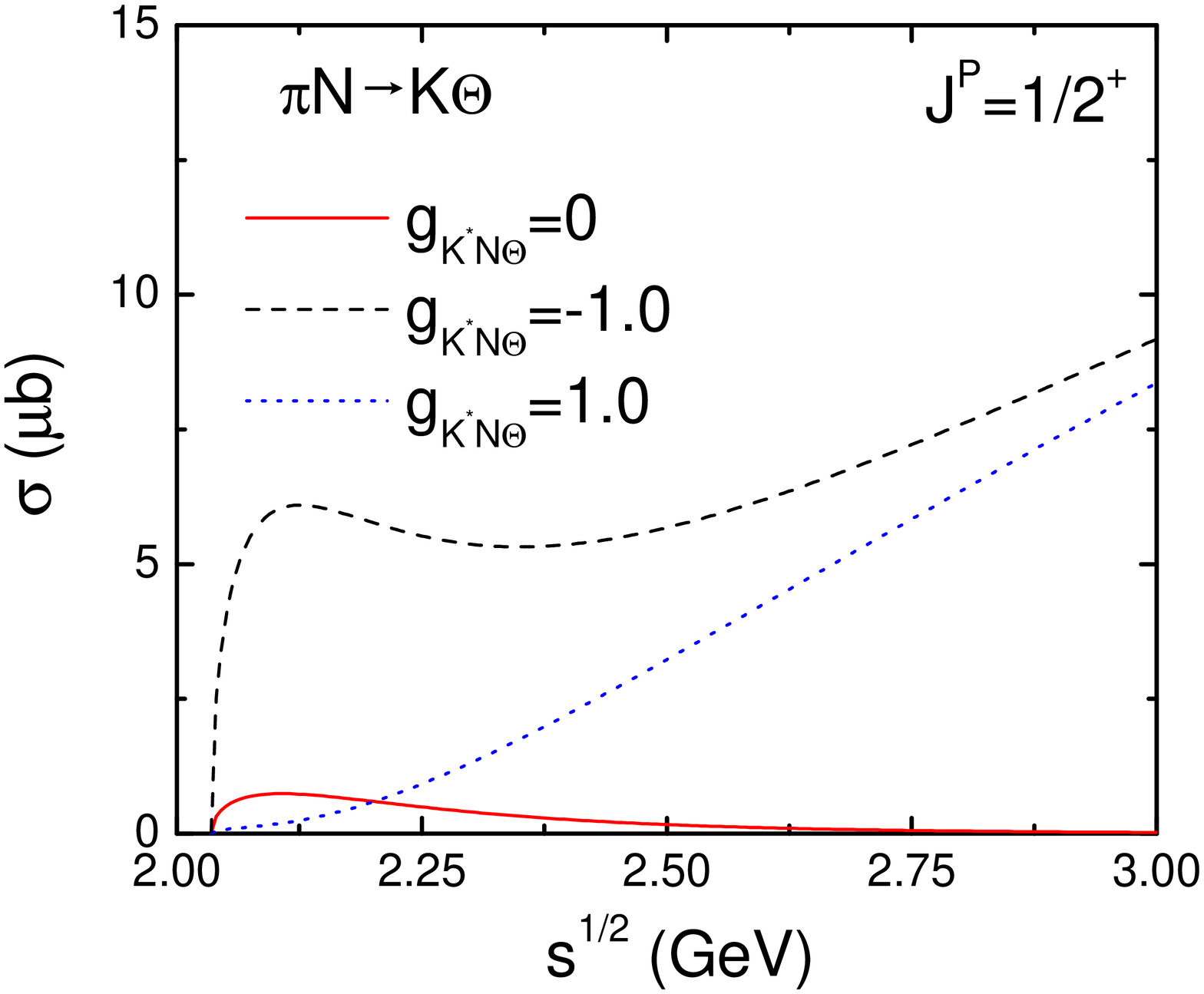}
\hspace{-0.77cm}
\includegraphics[height=3.3cm,width=4cm,angle=-90]{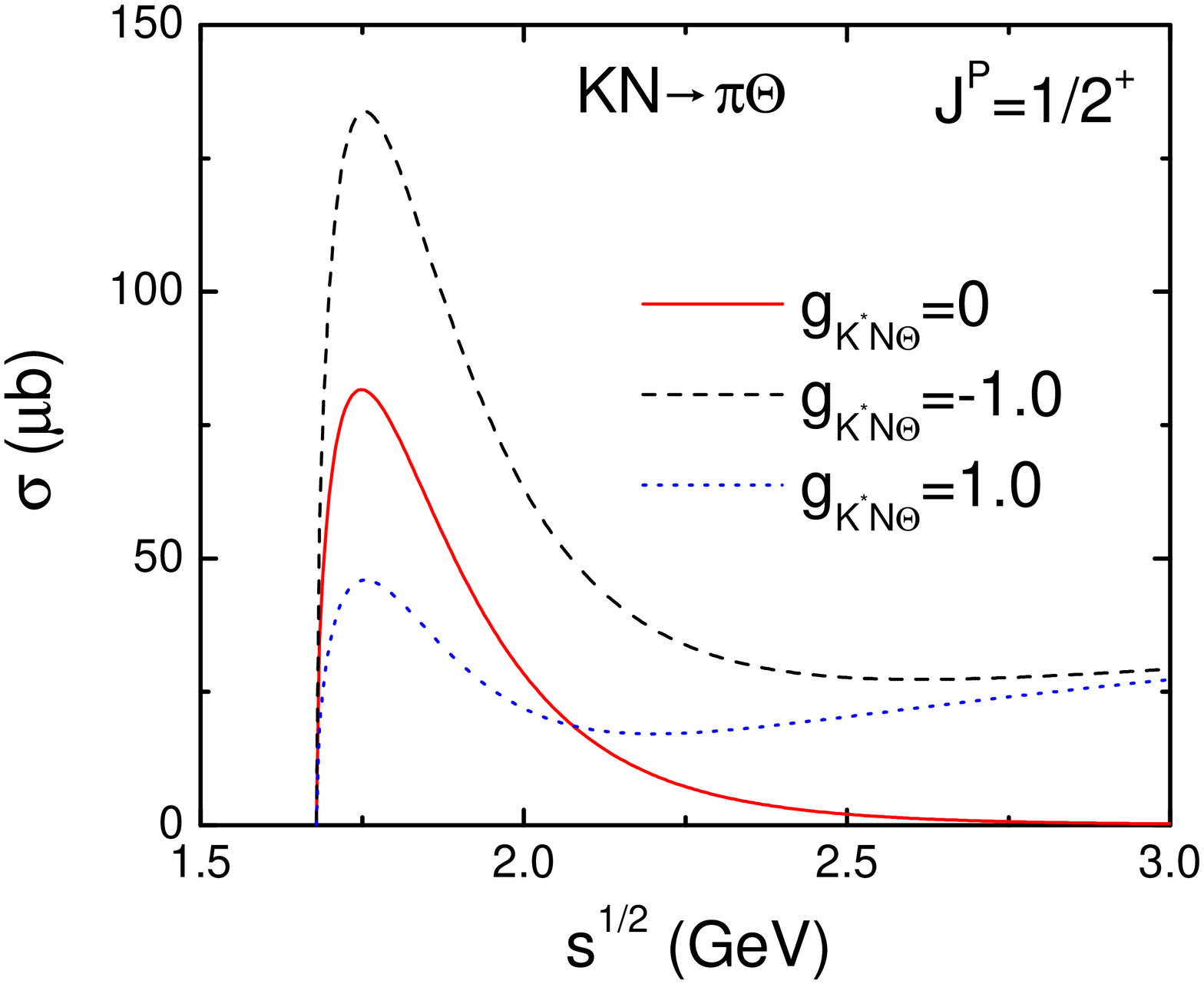}
\hspace{-0.77cm}
\includegraphics[height=3.3cm,width=4cm,angle=-90]{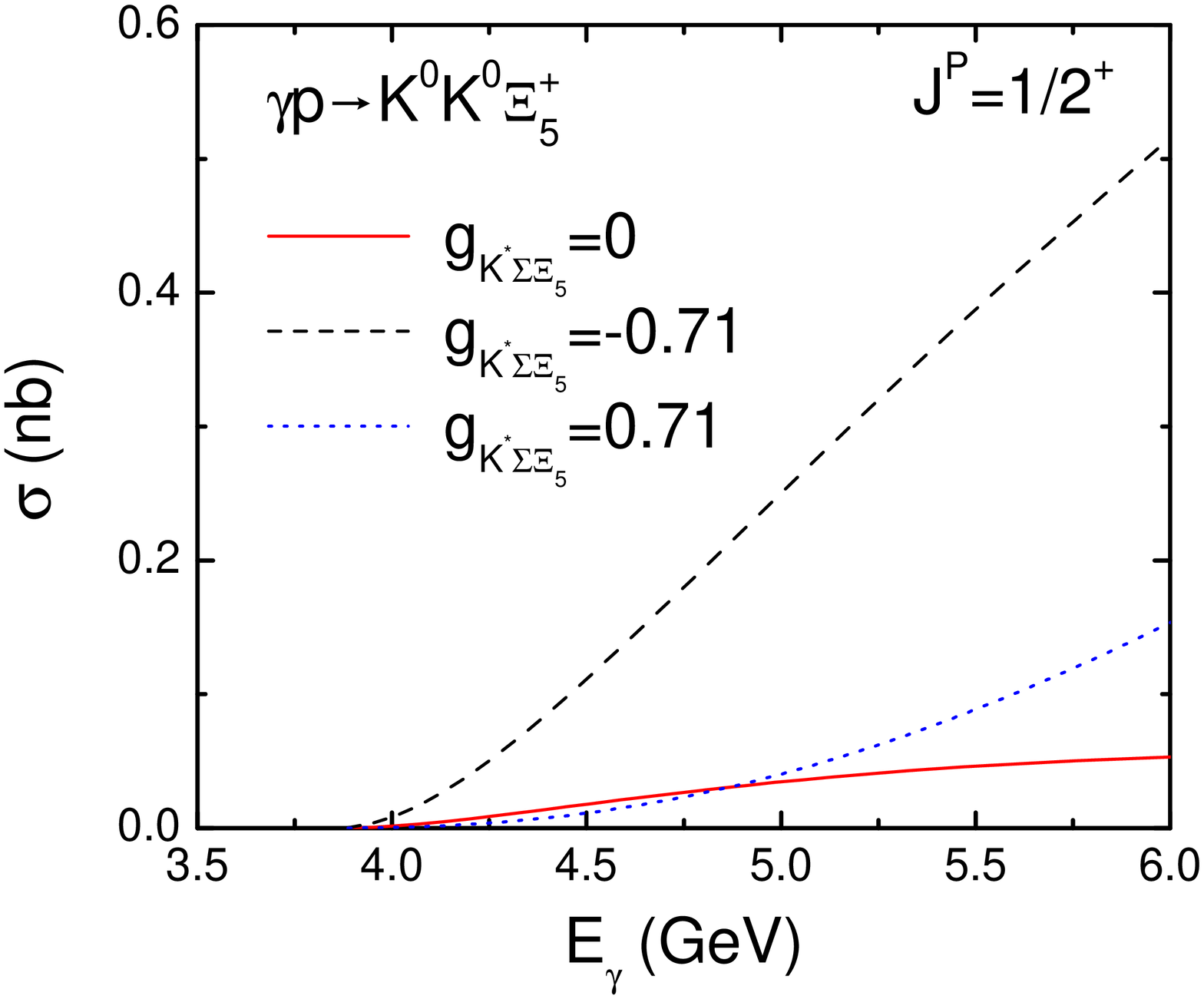}
\hspace{-0.77cm}
\includegraphics[height=3.3cm,width=4cm,angle=-90]{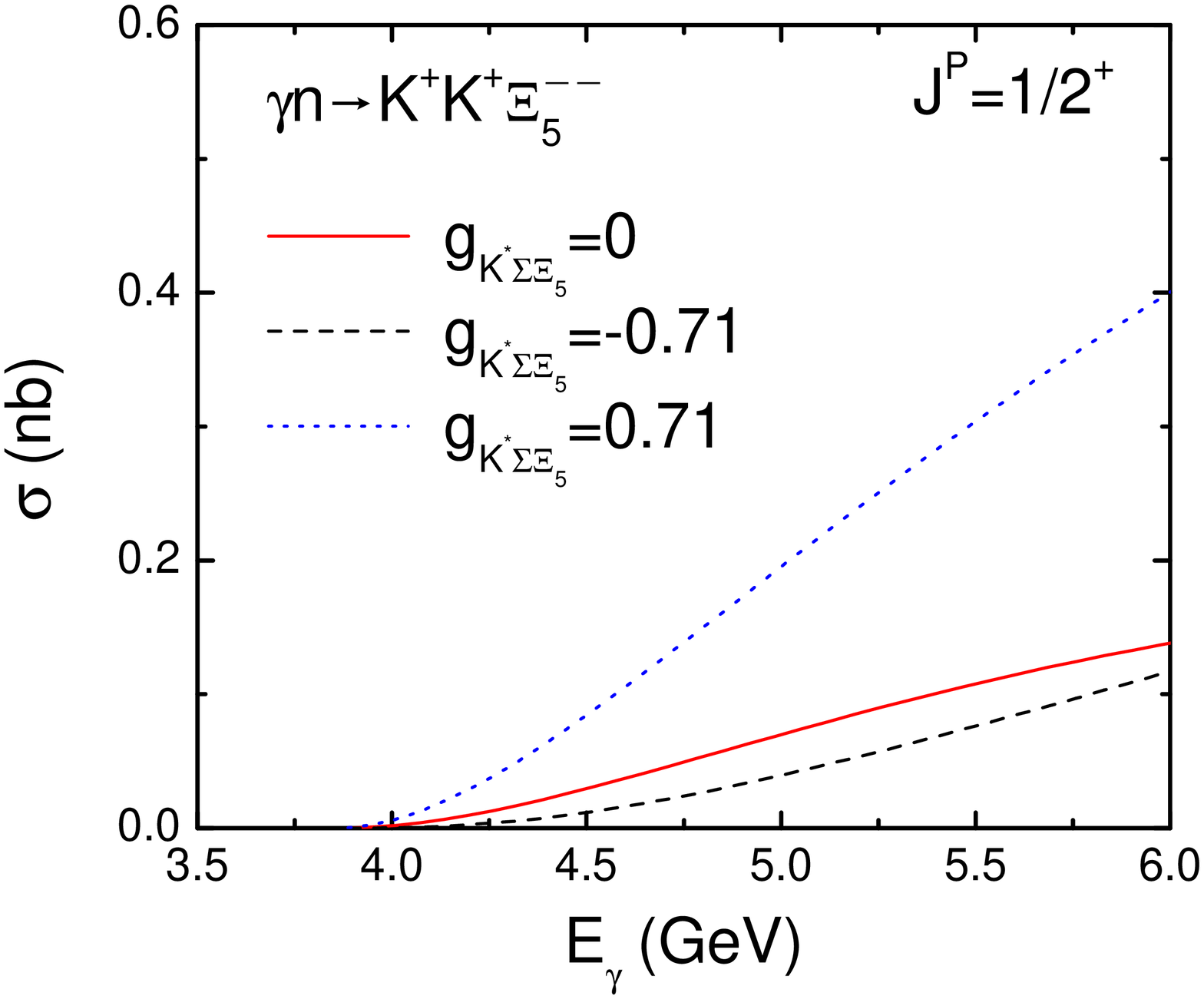}}
\caption{Cross sections for $\Theta^+$ production from the reactions
$\pi N\to\bar K\Theta^+$ and $KN\to\pi\Theta^+$ (left two panels)
and $\Xi^{--}_5$ and $\Xi_5^+$ from photon-nucleon reactions (right two
panels).}
\label{crosspik}
\end{figure}

For $\Theta^+$ production from the reactions $\pi N\to\bar K\Theta^+$
and $KN\to\pi\Theta^+$, the relevant diagrams are shown in 
Fig.\ref{diagrampik}. Using additional known empirical coupling
constant $g_{\pi KK^*}=3.28$ and similar form factors as in the reaction
$pp\to\Sigma^+\Theta^+$, their cross sections are shown in the left
two panels of Fig.\ref{crosspik} with values of 0.5-5 $\mu$b for 
$\pi N\to\bar K\Theta^+$ and 20-75 $\mu$b for $KN\to\pi\Theta^+$ at
center-of-mass energy of 0.2 GeV above their respective threshold
energies. It will be interesting to compare the predicted cross
sections with future data from experiments that are being carried out
at KEK in Japan.

\section{$\Xi_5^+$ and $\Xi_5^{--}$ production in photonucleon reactions}

\begin{figure}[ht]
\centerline{\includegraphics[height=2.5cm,width=11cm,angle=0]{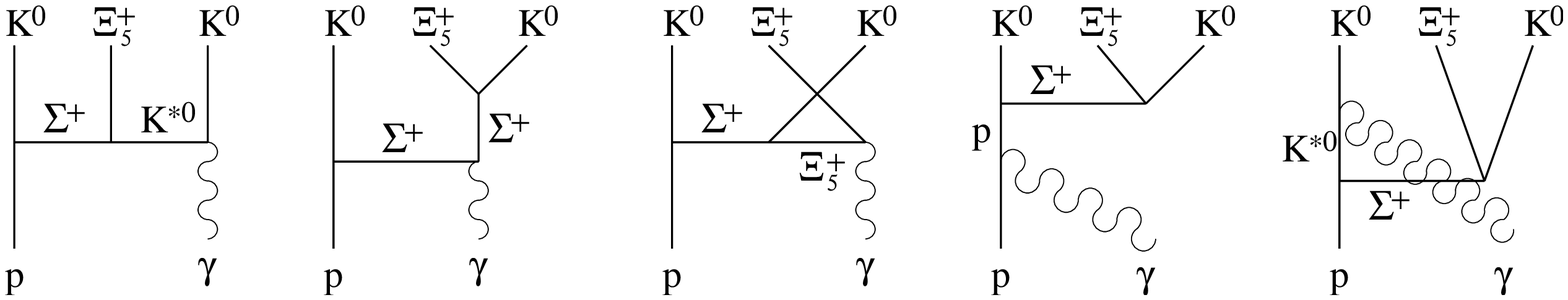}}
\caption{Diagrams for $\Xi^+_5$ production from the reaction
$\gamma p\to K^0K^0\Xi^+_5$.} 
\label{diagramx}
\end{figure}

For $\Xi_5^+$  production in photonucleon reactions, the relevant
diagrams are shown in Fig.\ref{diagramx}. The needed additional
coupling constants $g_{K(K^*)\Sigma\Xi_5}$ can be obtained from the SU(3)
relations $g_{K(K^*)\Sigma\Xi_5}=g_{K(K^*)N\Theta}/\sqrt{2}$ if
$\Theta^+$ and $\Xi_5$ belong to the same antidecuplet \cite{oh3}. 
Introducing appropriate form factors with empirical cutoff parameters 
\cite{liu3}, the resulting cross section for the reaction
$\gamma p\to K^0K^0\Xi^+_5$ together with that for the reaction
$\gamma n\to K^+K^+\Xi_5^{--}$ are shown in the right two panels of
Fig.\ref{crosspik} for $g_{K\Sigma\Xi_5}=0.71$ but different
values of $g_{K^*\Sigma\Xi_5}$. The cross sections are about 0.01-0.1
nb at photon energy $E_\gamma=4.5$ GeV, and this information will be 
useful for planning experiments to verify these exotic multistrange
pentaquark baryons in photonucleon reactions.

\section{$\Theta^+$ production in relativistic heavy ion collisions}

The quark-gluon plasma formed in heavy ion collisions at the
Relativistic Heavy Ion Collider (RHIC) \cite{qm} provides a promising
environment for producing hadrons consisting of multiple quarks such
as the pentaquark baryons. Based on the quark coalescence model for
hadron production from the quark-gluon plasma \cite{greco}, which has
been shown to give a consistent description of the observed large
baryon/meson ratio at intermediate transverse momenta of about 3 GeV/c
and scaling of hadron elliptic flows according to their constituent
quark content, we have evaluated the yield of $\Theta^+$ in Au+Au
collisions at $\sqrt{s_{NN}}=200$ GeV. With a plasma of quarks of
constituent masses and massive gluons at temperature $T_c=175$ MeV and
in a volume $\sim 1000$ fm$^3$ as in Ref.\cite{greco}, the resulting
$\Theta^+$ number is about 0.19 if $\Theta^+$ has a radius of $\sim
0.9$ fm, and is somewhat smaller than that predicted by the
statistical model \cite{randrup}. To take into account subsequent
hadronic effects due to the reactions 
$\Theta\leftrightarrow KN$, $\Theta\pi\leftrightarrow KN$, and 
$\Theta\bar K\leftrightarrow\pi N$, a schematic isentropic expanding
fireball model has been used \cite{chen}.

Using the cross sections evaluated in previous sections, the time
evolution of the $\Theta^+$ abundance has been studied for different
values of $\Theta^+$ width or coupling constant $g_{KN\Theta}$. The
final $\Theta^+$ yield increases to about 0.5 for $\Gamma_\Theta=20$
MeV but remain about 0.2 for $\Gamma_\Theta=1$ MeV. Similar effects
are seen if the number of $\Theta^+$ produced from the quark-gluon
plasma is increased to 0.94, resulting from a $\Theta^+$ radius 
that is 30\% larger, or decreased to zero. With the expected small
$\Theta^+$ width, the hadronic effect is thus small, and the abundance
of $\Theta^+$ is then sensitive to its production from the quark-gluon
plasma. Since the baryon chemical potential in the quark-gluon plasma
is small, the yield of anti-$\Theta$ is only slightly less than that
of $\Theta^+$.

\section{Summary}

We have studied exotic pentaquark baryon production both in elementary
reactions involving photons, protons, pions, and kaons on nucleon
targets and in relativistic heavy ion collisions. With a
phenomenological hadronic model, cross sections for the elementary
reactions are evaluated by taking into account the coupling of 
$\Theta^+$ to both $KN$ and $K^*N$. Assuming that $\Theta^+$ has
positive parity and a width of $1$ MeV, corresponding to a coupling
constant $g_{KN\Theta}\approx 1$, and depending on the value of
$g_{K^*N\Theta}$, the cross sections are 4-35 nb for 
$\gamma p\to\bar K^0\Theta^+$, 15-30 nb for $\gamma n\to K^-\Theta^+$, 
4-15 nb for $\gamma p\to\bar K^{*0}\Theta^+$, and 5-150 nb for 
$\gamma n\to K^{*-}\Theta^+$ at photon energy $E_\gamma=3$
GeV; 0.1-2 $\mu$b for $pp\to\Sigma^+\Theta^+$, 0.5-5 $\mu$b for $\pi
N\to\bar K\Theta^+$, and 20-75 $\mu$b for $KN\to\pi\Theta^+$ at
center-of-mass energy of 0.2 GeV above their threshold. The produced
$\Theta^+$ in these reactions all peaked at forward angles in the
center-of-mass system. The cross sections are smaller by about an
order of magnitude if $\Theta^+$ has negative parity. For multistrange
pentaquark baryons $\Xi_5^+$ and $\Xi_5^{--}$, the cross sections for
their production in the reactions $\gamma p\to K^0K^0\Xi^+_5$ and
$\gamma n\to K^+K^+\Xi^{--}_5$ are 0.01-0.1 nb at photon energy
$E_\gamma=4.5$ GeV. In heavy ion collisions at RHIC, about 0.19
$\Theta^+$ is produced during hadronization of the quark-gluon
plasma, but the final number after hadronic absorption and regeneration
depends on the width of $\Theta^+$. These results are useful
for understanding not only the experimental data but also the
properties of pentaquark baryons.

\vspace{-0.1cm}

\section*{Acknowledgments}

We would like to thank Lie-Wen Chen, Vincenzo Greco, and Su Houng Lee
for discussions and collaboration on some of the reported works.

\end{document}